\documentstyle[12pt]{article}
\input epsf
\begin{document}

\title{ON THE SUTHERLAND'S INTEGRABILITY CONDITION \\
       FOR TWO-DIMENSIONAL N-PARTICLE SYSTEMS. }

\author{A. AZHARI $^{\star, 1}$    \hskip1truecm and 
\hskip1truecm  T.T. TRUONG $^{\dagger, 2}$ \\
                                             \\
        $ \star $) Laboratoire de Mod\`eles de Physique Math\'ematique,\\
                      D\'epartement de Physique, Universit\'e de Tours,\\
                          Parc de Grandmont, F-37200 Tours, France. \\ \\
        $ \dagger $) Groupe de Physique Statistique. D\'epartement de Physique, \\                    Universit\'e de Cergy-Pontoise B.P.8428,\\                                        F-95806 Cergy-Pontoise Cedex France. }

\maketitle

\begin{abstract}

Following Sutherland's work on one-dimensional integrable systems we
 formulate and study its two-dimensional version.Physically it expresses the
 absence of true 3-body forces among an assembly of N particles 
leaving exclusively effective 2-body interactions.This criterion
 may be a suitable candidate for an  integrability condition. 
\end{abstract}

\bigskip
\noindent $^1${e-mail: azhari@celfi.phys.univ-tours.fr}\\
\noindent $^2${e-mail: truong@u.cergy.fr}
\pagebreak

\section{Introduction}

In the last twenty years, the quantum N-body problem has been
chiefly studied in one space dimension with great success thanks to the 
discovery of integrable systems. Among those, one can count systems soluble
with the Bethe ansatz wave function as well as others systems discovered 
originally by Calogero and Sutherland at the beginning of the 70's. The last
topic has evolved considerably  and connections to many algebraic 
structure have been established \cite{Ol}.\\

 One main ingredient to the solubility is the fact that the dynamics of
 the N-particle system is only due to pair potential and eventually
 external applied fields. This abscence of other type of N-body
 forces renders the problem completly soluble whether it is relativistic
 or not.\\

 Few breakthroughs have been made in 2 or 3 space dimensions. Due
 to more ``room'' available, particles can go around each 
other instead  of being confined on the line  where they must scatter 
inevitably. The resulting effect is that it is difficult to give a
 concept of ``integrable'' systems. Yet Calogero and Machioro \cite{Cal} have discovered 
that if one includes some type of 3-body forces which are highly dependent on 
direction, one may obtain exactly soluble multiparticle dynamics in
 three space dimensions. \\

 The interest in recent years in two-dimensional quantum systems
stems from the discovery of the fractional quantum Hall effect which
 indirectly revived an older work of Leinass and Myrheim \cite{lein} on quantum 
theory of many-particles in two-dimensions as well as quantum dots and electronic plasmas  \cite{Joh}. At the center of
 this topic are new objects, the anyons  
which seems to be responsible for some physical phenomenona (such as fractional
 quantum Hall effect)\cite{Wil} or realizations of ``exotic'' statistics which
 are not excluded in two-dimensions. Hence different many-particle
 wavefunctions 
 have been proposed in this spirit such as the known Laughlin wavefunction
 for the fractional quantum Hall effect \cite{Lau,Laug}.\\
 
In this paper we shall not discuss the physical phenomena, but
 instead pose the following problem: under which condition, in
 two-dimensions can one have only pair interactions? And more
 generally when does one
 obtain an integrable system? We shall see then that the emergence
 of fractional statistics appears to be natural and consistent with
 the existence of pair-potentials among the N-particles, only for a specific
 class of pair-potentials.\\            

Let us first recall that in non-relativistic quantum mechanics of a single
    particle the wavefunction depends on two real variables $x$   and  $y$ i.e. $\psi(x,y)$.
One may equivalently use the complex combinations:
\begin{equation}
     z=x+iy \hskip2truecm {\rm and} \hskip2truecm \overline{z}=x-iy
\end{equation}
and consider instead $\psi(z,\overline{z})$;  this last wave function is in fact a 
restriction of a fonction of two complex variables $\psi(z,z')$
, such that $z'=\overline{z}$. To account for manifest correlations
 among particles of the systems it is natural to postulate the Bijl-Dingle-Jastrow
 wavefunction \cite{SuA4, Bijl} which is generally proposed  in one-dimension to describe
 the ground state of a system of N particles.\\
\begin{equation}     \Psi=\Psi(z_{1},\overline{z}_{1};\ldots ;z_{N},\overline{z}_{N})
              =\prod_{\scriptstyle i<j}\psi(z_{ij},\overline{z}_{ij})
\end{equation}
  where $z_{ij}=z_{i}-z_{j}$. Excited states may be constructed from the ground state using a standard method \cite{SuA5}.\\

In this wavefunction, the order of  pairs of particles is single
 out by the pair wavefunction  $\psi(z_{ij},\overline{z}_{ij})$. This is a
 general feature in all soluble N-particle systems in one-dimension, and
 becomes thus a valuable starting point for the study
 of two-dimensional systems.\\

In section~I, we review the situation in one-dimension to establish
the grounds for such procedure. There, the central object is the Sutherland's
condition for the solubility of the problem. This condition merely states 
 that the 3-body potential arising from a state described by the 
 Bijl-Dingle-Jastrow wavefunction, may be recast into a sum of
 pair-potentials so that the whole system behaves pratically under pure 
effective pair potentials. This is our main argument and this may
 be related to what is known in integrable quantum field theory in
 one-space dimension. In fact when the S-matrix of such a quantum field 
theory is factorizable or can be written as product of S-matrices for
 pairs of particles, subjected to the usual conditions of unitarity and 
 analyticity, does one have an infinite number of concerved quantities
 in the theory. What is more interesting is that A.B. Zamolodchikov and  Al.B.
Zamolodchikov \cite{Zam} have shown that in the non-relativistic limit, these
 pair S-matrices reproduce the scattering phase shifts of the soluble
 pair potentials in one-dimension. It is then tempting to identify
 the Sutherland's
 condition as the integrability condition in one-dimension. In fact this 
is really so because such a condition  contains all the known soluble pair 
potentials.\\

Our objective is thus to explore the Sutherland's condition in
 two-dimensions. In section~II, we shall derive the condition of
 Sutherland in two dimensions 
and study its properties in details. As we shall see, many new aspects
emerge as compared to those in one-dimension: the inclusion of fractional 
statistics, the nature of the effective pair-potentials, and the triviality
of the attractive harmonic potential in two-dimensions.\\

In section~III, we shall raise the question whether there exists a repulsive
interaction among pairs which might insure stability  against the
 collapse due to the harmonic attractive  potential . In fact we shall
 present a two-dimensional version of the Sutherland elliptic potential
as a concrete example. The model has the merit of providing an example 
of Wigner solid in two dimensions. In the scaling limit of this potential
we shall see how the fact that particles can never ``over take'' each
 other in two-dimensions is the key to the solubility of the problem. In fact the particles are locked inside two-dimensional rectangular cells and remain 
``impenetrable''.\\

We conclude by presenting some ideas on how one may generalize the previous 
ideas to find non-trivial integrable systems in two-dimensions as well 
as to seek limits to construct integrable relativistic quantum field
 theories in two-space dimensions.
\section{The one-dimensionnal case revisited}.

This has been worked out in the seventies by B.Sutherland \cite{SuA4,SuA5} and F.Calogero \cite{Cal1,Cal2}.
 The N-particule wavefunction is assumed to be of the Dingle-Bijl-Jastrow
 form:\\
\begin{equation}
\Psi=\prod_{\scriptstyle i<j}|\psi(x_{i}-x_{j})|^\lambda
\end{equation}
where $\psi(x)$ is the pair wavefunction and $\lambda$ a real constant.\\

Introducing the associated functions $\phi$ and $\varphi$ by:
\begin{equation}
\psi=\exp{\phi}  \hskip4.5truecm            \varphi=\frac{d\phi}{dx}
\end{equation}
and applying the kinetic energy operator on $\Psi$, one obtains:
\begin{equation}
\begin{array}{lr}
-\displaystyle\sum_{i=1}^N\frac{\partial^2}
{\partial x_{i}^2}\Psi=\{-2 \lambda \sum_{
\scriptstyle j<i}[\varphi'(x_{i}-x_{j})+
\lambda \varphi^2(x_{i}-x_{j})] & \\ 
\hskip2.7truecm  +2 \lambda^2 \sum_{\scriptstyle i,j,k}[\varphi(x_{k}-x_{i})\varphi(x_{i}
-x_{j})+\varphi(x_{i}-x_{j})\varphi(x_{j}-x_{k})\\
\hskip2.7truecm   +\varphi(x_{j}-x_{k})\varphi(x_{k}-x_{i})]\}\Psi. & \\
\end{array}
\end{equation}

The first term of the r.h.s. of (5) represents the pair potentiel 
between pairs of particles due to the choice of $\psi(x)$.
 the second term is an induced 3-body potential beetwen any triplet
i,j,k of the N-particles.\\

 Sutherland proposed then to choose $\psi$ such that this 3-body
 potential breaks up into additional pair potentials, namely that:
\begin{equation}
\varphi(x)\varphi(y)+\varphi(y)\varphi(z)+\varphi(z)\varphi(x)=
 f(x)+f(y)+f(z)
\end{equation}
for\hskip5truecm   $x+y+z=0$ \\

The form of $f(x)$  being essentially due to the choice of $\psi(x)$.
 If such a choice is made, then there exists among
 the N-particles only an effective pair potential:
\begin{equation}
V(x)=\lambda[\varphi'(x)+\lambda\varphi^2(x)-\lambda f(x)].
\end{equation}

The philosophy of this statement is analoguous to the one
 adopted in the theory of integrable quantized fields in
 1+1-dimensions. There it is stated that the N-body S-matrices  are
 factorized into two-body S-matrices. In fact Zamolodchikov and
 Zamolodchikov \cite{Zam} have shown that in one of these theories, the
 non-relativistic limit of such 2-body S-matrix is precisely the phase-shift in a relative pair potential of the type (7). In this sense,
 one may say that the Sutherland's condition (6), is in fact an 
 integrability condition for an N-body problem in one-dimension.\\

To get more insight we may advantageously replaced the condition (6) 
by the following one, using elementary algebra:
\begin{equation}
(\varphi(x)+\varphi(y)+\varphi(z))^2=g(x)+g(y)+g(z)
\end{equation}
with\hskip3truecm  $x+y+z=0$\hskip1truecm and \hskip1truecm  
$g(x)=\varphi^2(x)+2f(x)$.\\

The form of (8) turns out to be exactly a relation satisfied by the 
Weierstrassian elliptic functions $\zeta(x)$ and P(x) namely:
\begin{equation}
(\zeta(x)+\zeta(y)+\zeta(z))^2=P(x)+P(y)+P(z)
\end{equation}
with \hskip4truecm $x+y+z=0$ \\
Sutherland who discovered this connection identified then $\psi(x)$ to
 $\sigma(x)$, the Weierstrassian $\sigma$-function . The periodicity of 
$\sigma(x)$ is instrumental in exhibiting an example of Wigner
 solid in one-dimension  \cite{Su35,Su34}.\\

Some remarks on the properties of $\varphi(x)$ are now in order:\\

a) $\varphi(x)$ is in fact defined up to a linear term: the substitution 
$\varphi(x) \to \varphi(x)+ax+b$ leaves relation (6) invariant.The linear
term $ax+b$ induces a pair potential:
\begin{equation}
V(x)=\lambda\{a+\frac{\lambda}{2}(ax+b)^2-\lambda x\}
\end{equation}
which is essentially a shifted harmonic oscillator potential.
 In this sense, the harmonic oscillator pair potential is simply
 a trivial one.\\

b) As shown by Sutherland \cite{SuA4}, particular limits of the $\zeta(x)$
 potential reproduce all the non-singular pair-potential known
 in one-dimension.\\

c) We note also that the pair $\delta$-function
 is also contained  in (6) which in this case the r.h.s of \cite{Su34}
 is simply constant
  and the $\Psi$ take up the form of a Bethe ansatz wavefunction.
\section{Two-dimensional case}

As pointed out before, the pair wavefunction is in fact a restricted 
wavefunction of 2 complex variables $\psi(z,z')$ with $z'=\overline{z}$.
 In analogy to  section 2 we shall introduce the notations:
\begin{equation}
\begin{array}{lll}
\psi&=&\exp{\phi} \\
 \varphi&=&{\displaystyle\frac{d\phi}{dz}}\\
\end{array}\hskip3truecm
\begin{array}{rrc}
\phi&=&\phi(z,\overline{z})\\
\overline{\varphi}&=&{\displaystyle\frac{d{\phi}}{d\overline{z}}}\\ 
\end{array}
\end{equation}
The application of the kinetic energy operator, assuming that particles are
 of unit mass.\\
$$\sum_{i=1}^{N}\frac{\partial^2}{\partial\overline{z}_j\partial z_i}=-\frac{1}
{2}\sum_{i=1}^{N}(\frac{\partial^2}{\partial x_{i}^2}+\frac{\partial^2}{\partial y_{i}^2})$$
on a N-particle Dingle-Bijl-Jastrow wavefunction yields a sum
 of pair-potentials and an induced 3-body potential as in one-dimension.
Therefore following Sutherland we are led to  the generalized condition:
\begin{equation}
\begin{array}{lr}
\hskip1truecm \{\hskip.35truecm \varphi(x,\overline{x})\overline{\varphi}(y,\overline{y}) \hskip0.5truecm +\hskip0.5truecm
  \varphi(y,\overline{y})\overline{\varphi}(x,\overline{x}) & \\
\hskip1truecm +\hskip0.3truecm \varphi(y,\overline{y})\overline{\varphi}(z,\overline{z})\hskip0.5truecm +\hskip0.5truecm
  \varphi(z,\overline{z})\overline{\varphi}(y,\overline{y}) & \\
\hskip1truecm +\hskip0.3truecm \varphi(z,\overline{z})\overline{\varphi}(x,\overline{x})\hskip0.5truecm +\hskip0.5truecm 
  \varphi(x,\overline{x})\overline{\varphi}(z.\overline{z})\hskip.35truecm \} & \\ 
\hskip4truecm = [ f(x,\overline{x}) + f(y,\overline{y}) + f(z,\overline{z}) ]& \\
\end{array}
\end{equation}
\hskip1truecm with:$$ x+y+z=0\hskip1truecm  {\rm and}\hskip1truecm \overline{x}+\overline{y}+\overline{z}=0 $$ which evidently states that the 3-body potential between any triplet of particles will be break up into a sum of 2-body potentials.\\

Unfortunatly, there is up to now no theory of factorized S-matrices
 in (2+1) dimensions. Yet, we may call this condition an integrability
 condition if non-trivial interesting solutions can be found. In the sequel we shall seek to construct some solutions based on experience in one-dimension.\\

But before doing so let us mention that an alternative way for formulating (12) would be, in analogy to eq (8)
\begin{equation}
\begin{array}{lr}
\{\varphi(x,\overline{x})+\varphi(y,\overline{y})+\varphi(z,
\overline{z})\}\{\overline{\varphi}(x, \overline{x})+\overline{\varphi}
(y,\overline{y})+\overline{\varphi}(z,\overline{z})\} & \\
\hskip6truecm  =g(x,\overline{x})+g(y,\overline{y})+g(z,\overline{z}) & \\
\end{array}
\end{equation}
whenever:$$ x+y+z=0\hskip1truecm {\rm and}\hskip1truecm \overline{x}+
\overline{y}+\overline{z}=0 $$

Under this form, eq.(13) remains obviously invariant under the
 double substitution:
\begin{eqnarray}
\varphi(x,\overline{x})\to\varphi(x,\overline{x})+ax+a'\overline{x}+b & \\
\overline{\varphi}(x,\overline{x})\to\overline{\varphi}(x,\overline{x})
+\overline{a'}x+\overline{a}\overline{x}+\overline{b} 
\end{eqnarray}
Again this linear part is responsible for the shifted harmonic oscillator
pair-potential between particles. Thus such a potential is of trivial
nature and has tendency to cause a collapse to the center
 of mass of the system of N-particles. Morever,$$\phi \to
\phi+a'z\overline{z}
+\frac{1}{2}(az^2+\overline{a}\overline{z}^2)
+bz+\overline{b}\overline{z}+{\rm const}.$$
and the new potential is:
\begin{equation}
V\to(V+a')+\varphi(x,\overline{x})+\overline{\varphi}(x,
\overline{x})+(b+\overline{b}).
\end{equation}
This transfomation may be used  to generate new pair potential
 from a known one.The two-dimensional effective pair-potential in general
 has the form:\\
$$V=\lambda(\lambda - 1)\frac{\partial \phi}{\partial z} \frac{\partial \phi}
{\partial \overline{z}} + \lambda ^2  \frac{\partial^2 \phi}{\partial z \partial 
\overline{z}} - \frac{\lambda ^2}{2}f(z, \overline{z})$$  
We are now in a position to study some particular situations:\\

a) If $\phi$ is a solution of the Laplace operator\\
$$\frac{\partial^2}{\partial z\partial\overline{z}}\phi=0$$
Then $\phi=f(z)+g(\overline{z})$, here $ f $ and $g$ are independent
 functions not related to those of eq.(12) or (13). Sutherland's condition
 becomes
\begin{equation}
\{f'(x)+f'(y)+f'(z)\}\{g'(\overline{x})+g'(\overline{y})+g'
(\overline{z})\}=h(x,\overline{x})+h(y,\overline{y})+h(z,\overline{z})
\end{equation}
\hskip2truecm for $$x+y+z=0.$$
We note that if $g'=$const ( or resp. $f'=$const ) the condition (17) is
 automatically satisfied: $\psi=\exp{f(z)}$ or $\psi=\exp
 {g(\overline{z})}$ is a solution.\\

b) If $ \phi=\phi(z\overline{z})$, function of the distance $ z\overline{z}$, this represents a physically reasonable situation for which 
the Sutherland 's condition is:
\begin{equation} 
\begin{array}{lr}
\phi'(y\overline{y})\phi'(x\overline{x})(\overline{x}y+\overline{y}x)+
\phi'(y\overline{y})\phi'(z\overline{z})(y\overline{z}+\overline{y}z) 
+ \phi'(z\overline{z})\phi'(x\overline{x})
(x\overline{z}+z\overline{x}) & \\
\hskip5truecm  = h(x,\overline{x})+h(y,\overline{y})+h(z,\overline{z}) 
\end{array}
\end{equation}
Note that for $\phi'=$const one recovers the harmonic oscillator
 pair-potential, which has been recently investigated by Mushkevich et al \cite{Mus}. We observe that if the pair wavefunction is taken az $\psi(z, \overline
{z})\cong z^\alpha \exp{(-z\overline{z})}$  for example, the effective pair-potential is\\
$$ V=(\frac{3}{2}\lambda ^2 - \lambda)z\overline{z}-\lambda + \alpha(\frac{\lambda ^2}{2}
-\lambda)$$
Thus the anyonic factor $ z^{\alpha} $ simply shifts $ V $ by a constant amount. In fact this is the only known instance where exotic statistics seem to be consistent
with the Sutherland condition.\\ 

Since:$$\overline{x}y+\overline{y}x=2\vec{x}.\vec{y}$$
represents the scalar product of the vector $\vec{x}$ and $\vec{y}$ in the plane, 
the Sutherland's condition takes a new vector form as a scalar product:
\begin{equation}
[\phi'(|\vec{x}|)\vec{x}+\phi'(|\vec{y}|)\vec{y}+\phi'
(|\vec{z}|)\vec{z}]^2=h(\vec{x})+h(\vec{y})+h(\vec{z})
\end{equation}
\hskip1.5truecm with  $$ \vec{x}+\vec{y}+\vec{z}=\vec{0}$$

This has the same structure as eq.(6). Calogero and Machioro \cite{Cal}   have treated in three-dimensions, in the
 same spirit, the problem of N-particles only with potential
 dependent on the interparticle distance. In other words the wave
 function is of the Dingle-Bijl-Jastrow type. However they kept the
 induced three-body potential and have not look at the
possibility it may decompose into pair-potentials.
\section{Two dimensional Sutherland's model}

Inspired by the work of Sutherland we may extend his model in two-dimensions
 obtaining a Wigner solid in the plane. The wavefunction $\Psi$ is taken 
as :
\begin {equation}
\Psi=c\prod_{\scriptstyle i<j}\Theta_1(\frac{\Pi}{L}(x_i-x_j))
\Theta_1(\frac{\Pi}{L'}(y_i-y_j))
\end{equation}
where $\Theta_1$ is the Jacobian odd-theta function and $L$, $L'$ are lengths in the x and y directions.

Applying the kinetic energy operator on the wavefunction $\Psi$ yields :

\begin{equation}
\begin{array}{lr}
\frac{1}{\displaystyle\Psi}(\displaystyle\sum_{i=1}^N(\frac{\partial^2}{\partial x_i^2
}+\frac{\partial^2}{\partial y_i^2})\Psi) = 
 \displaystyle\sum_{\scriptstyle i<j}[N\frac
{\Theta''_1(\displaystyle\frac{\Pi}{L}(x_i-x_j);\frac{ir}{L})}{\Theta_1(\displaystyle\frac{\Pi}{L}
(x_i-x_j);\frac{ir}{L})}+2\zeta(\frac{L}{2})(N-2)\frac{1}{L}] & \\[0.8truecm]
\hskip3.8truecm +\hskip.15truecm\displaystyle\sum_{\scriptstyle i<j}
[N\frac{\Theta''_1(\displaystyle\frac{\Pi}{L}(y_i-y_j);\frac{ir'}{L'})}
{\Theta_1(\displaystyle\frac{\Pi}{L'}(y_i-y_j);\frac{ir'}{L'})}+2\zeta(\frac{L'}{2})(
N-2)\frac{1}{L'}] & \\
\end{array}
\end{equation}

In this case the integrability condition (13) will appears as being derived by a combination of (9).In fact we have :
\begin{equation}
\Phi= \ln \Theta_1(\frac{\Pi}{L}x)+\ln \Theta_1(\frac{\Pi}{L'}x)
\end{equation}
consequently
\begin{equation}
\varphi=\displaystyle\frac{d\Phi}{dz}=\frac{1}{2}\frac{d\Phi}{dx}+
\frac{1}{2i}\frac{d\Phi}{dy}
\end{equation}
or 
\begin{equation}
\varphi=\frac{1}{2}(\zeta(x)-\zeta(w)\frac{x}{w})-\frac{i}{2}(\zeta(y)-\zeta(w')\frac{y}{w'})
\end{equation}
where w and $w'$ are  the periods of the $\zeta(z)$-function of Weierstrass.\\

Similarly we obtain
\begin{equation}
\overline{\varphi}(z,\overline{z})=\frac{1}{2}(\zeta(x)-\zeta(w)\frac{x}{w})+\frac{i}{2}(\zeta(y)-\zeta(w')\frac{y}{w'})
\end{equation}

As observed before the parts in $\varphi$ and $\overline{\varphi}$ linear
 in $x$, $y$ and constants will not affect the integrability condition 
 the l.h.s. of (13) which in fact becomes 
\begin{equation}
\begin{array}{lcr}
\{[\zeta(x_1)+\zeta(x_2)+\zeta(x_3)]-i[\zeta(y_1)+\zeta(y_2)+\zeta(y_3)]\}
& \\
\{[\zeta(x_1)+\zeta(x_2)+\zeta(x_3)]+i[\zeta(y_1)+\zeta(y_2)+\zeta(y_3)]\}
 & \\
 = [\zeta(x_1)+\zeta(x_2)+\zeta(x_3)]^2+[\zeta(y_1)+
\zeta(y_2)+\zeta(y_3)]^2
\end{array}
\end{equation}
But this is precisely, because of the identity \cite{Ols}
$$[P(x_1)+P(x_2)+P(x_3)]+[P(y_1)+P(y_2)+P(y_3)] $$
if\hskip3truecm $x_1+x_2+x_3=0$\hskip1.5truecm $y_1+y_2+y_3=0$ \\

One may thus says that the  double Sutherland's model fulfills an integrability condition which is merely the sum of the separate integrability 
conditions in the $x$ and the $y$ directions.\\

To get a further insight on the integrability of this two dimensional problem one may consider one of its limiting case where the pair potential
 is essentially
\begin{equation}
V(\vec{x}-\vec{x'})=\frac{g}{(x-x')^2}+\frac{g}{(y-y')^2}
\end{equation}
Here $g$ is a coupling constant.\\

There one see that, relative to particle $\vec{x}$, particle $\vec{x'}$ 
remains constantly  in one of the quadrants centred at point 
$\vec{x}$, and conversely, this is the two-dimensional form of the non-
overtaking aspect of the dynamics a particle can never get out of the quadrant in which it is located with respect to its neighbor.(see fig.1)

\begin{figure}[thp]
\epsfxsize=12truecm
\epsfysize=7truecm
\epsffile{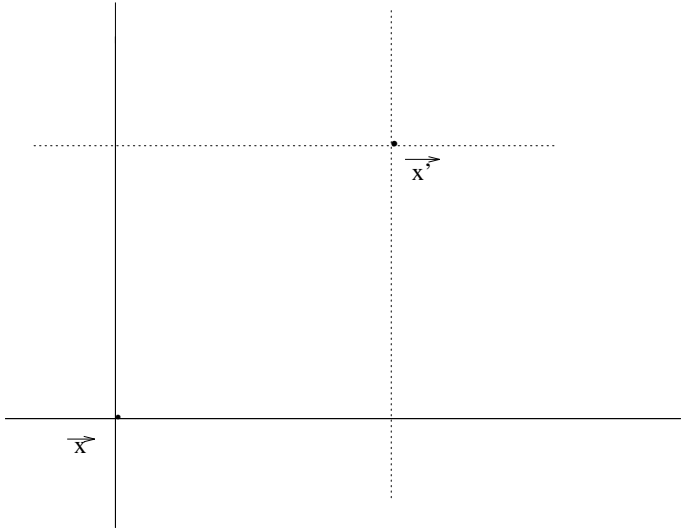}
\caption[]{
Particle $\vec{x}$ is in third quadrant of particle 
$\vec{x'}$ and particle
$\vec{x'}$ is in the first quadrant of
 particle $\vec{x}$. }
\end{figure}
Thus after a scattering the two particles will fly away from each
 other but continue to remain in these quadrant sectors for ever.\\

We can generalise this picture to an ensemble of N-particles
 and understand why these conservation laws are behind the integrability  of the model.\\

Although the periodical aspect leads to a two dimensional Wigner
 crystal, the particle dynamics in the particular limit of the pair-
potential is rather artificial due to the presence of ''forbidden lines''
 parallel to the axis attached to each particle.
\section{Conclusion and outlook.}
 
In this article we have tried to find a generalized version of the condition
found by Sutherland,for one dimensional integrable systems. Such systems 
admits  for ground state wavefunction as an N-particle wavefunction of 
the Dingle-Bijl-Jastrow form.\\

The Sutherland's condition merely states that for such a system there exists no true three-body potential but only effective pair-potentials.
 Sutherland was able to find the most general solution in one dimension,
it is a periodic one and its describes a Wigner crystal. In two dimensions the generalized Sutherland's condition does not admit obvious repulsive pair-potentials as in one dimension nor local $\delta$-function pair-potentials. But it does not exclude also other type of pair-potentials which remain to be discovered. We have constructed a simple example of two-dimensional solution fulfilling the generalized Sutherland's condition, this study
seems to suggest that a more general solution should be given by a 
2-variable theta functions depending on three modulu. Work is in progres
 in this direction. It is expected that special limits would lead to
 new form of soluble pair-potentials in two-dimensions.

\vfill\eject

\end{document}